\documentclass[pra,twocolumn,showpacs,superscriptaddress,10pt]{revtex4-1}
\usepackage[colorlinks,citecolor=blue,linkcolor=red,dvipdfm]{hyperref}
\usepackage{amsmath,amssymb,graphicx}
\usepackage{times}

\begin{document}
\title{One-dimensional gap solitons in quintic and cubic-quintic Fractional nonlinear Schr\"{o}dinger equations with a periodically modulated linear potential}

\author{Liangwei Zeng}
\affiliation{State Key Laboratory of Transient Optics and Photonics, Xi'an
Institute of Optics and Precision Mechanics of Chinese Academy of Sciences, Xi'an 710119, China}
\affiliation{University of Chinese Academy of Sciences, Beijing 100049, China}

\author{Jianhua Zeng}
\email{\underline{zengjh@opt.ac.cn}}
\affiliation{State Key Laboratory of Transient Optics and Photonics, Xi'an
Institute of Optics and Precision Mechanics of Chinese Academy of Sciences, Xi'an 710119, China}
\affiliation{University of Chinese Academy of Sciences, Beijing 100049, China}

\begin{abstract}
Competing nonlinearities, such as the cubic (Kerr) and quintic nonlinear terms whose strengths are of opposite signs (the coefficients in front of the nonlinearities), exist in various physical media (in particular, in optical and matter-wave media). A benign competition between self-focusing cubic and self-defocusing quintic nonlinear nonlinearities (known as cubic-quintic model) plays an important role in creating and stabilizing the self-trapping of D-dimensional localized structures, in the contexts of standard nonlinear Schr\"{o}dinger equation. We incorporate an external periodic potential (linear lattice) into this model and extend it to the space-fractional scenario that begins to surface in very recent years---the nonlinear fractional Schr\"{o}dinger equation (NLFSE), therefore obtaining the cubic-quintic or the purely quintic NLFSE, and investigate the propagation and stability properties of self-trapped modes therein. Two types of one-dimensional localized gap modes are found, including the fundamental and dipole-mode gap solitons. Employing the techniques based on the linear-stability analysis and direct numerical simulations, we get the stability regions of all the localized modes; and particularly, the anti-Vakhitov-Kolokolov criterion applies for the stable portions of soliton families generated in the frameworks of quintic-only nonlinearity and competing cubic-quintic nonlinear terms.
\keywords{Fractional calculus \and Cubic-quintic nonlinearity \and Nonlinear Schr\"{o}dinger equation \and Gap solitons}
\end{abstract}

\maketitle

\section{Introduction}
\label{sec1}

Standard quantum mechanics demands that every physically measurable quantity has always to be real and thus the resulting eigenvalues of every physical (quantum) operator should also be real. Such demand can only be satisfied if every physical observable is only connected with a Hermitian operator. However, studies, made in last decade, found that the standard quantum mechanics can be extended to non-Hermitian system which---counterintuitively---exhibits entirely real spectrum as well, supposing that the associated pseudo-Hermitian Hamiltonian is parity-time ($\mathcal{PT}$) symmetric \cite{Bender-PT1,Bender-PT2,Bender-PT3}. Advances of diverse kinds of solitons \cite{PT-our} and the propagation of linear and nonlinear waves in different $\mathcal{PT}$ -symmetric physical systems are abundant in recent years \cite{PT-rev1,PT-rev2,PT-rev3}. Another extension of the standard quantum mechanics was also done in the last decade: Laskin uncovered, from his several pioneering works \cite{Lask1,Lask2}, that the space-fractional quantum mechanics (SFQM) can arise in replacing the Brownian trajectories in Feynman path integrals with the L\'{e}vy flights. The fundamental physical model underpins the SFQM is the fractional Schr\"{o}dinger equation---the term was also coined by Laskin \cite{Lask2,Lask3}---which may be implemented in quantum physics \cite{Lask1,Lask2,Frac1} and  condensed-matter physics \cite{Frac-condensed}.

Particularly significant progress on the realization of fractional Schr\"{o}dinger equation (FSE) in optics has been made by Longhi in 2015  \cite{Frac-OL} based on transverse laser dynamics in aspherical optical cavities, opening a new path to study the properties and emergent phenomena of physical model involving fractional-diffraction. Since then, interesting results on generating and manipulating linear and nonlinear propagation dynamics of laser beams in such fractional optical models were obtained. Some typical works include:  Gaussian beams either evolved into diffraction-free beams \cite{Frac2} or undergone conical diffraction \cite{Frac6} during propagation without a potential, $\mathcal{PT}$ symmetry \cite{Frac3} and propagation dynamics of the super-Gaussian beams \cite{Frac7} , optical beams propagation with a harmonic potential \cite{Frac2,Frac6,Frac7} (which supports spatiotemporal accessible solitons too \cite{accessible1,accessible2}) and periodic potentials \cite{Frac3,Frac8}, propagation management of light beams in a double-barrier potential \cite{Frac9},  in the context of linear FSE regime; and in terms of nonlinear fractional Schr\"{o}dinger equation (NLFSE) regime \cite{Frac10,Frac13,Frac14,Frac5,Frac11,Frac12,Frac-NL}, including optical solitons (or solitary waves) without external potential \cite{Frac13,Frac14}, solitons supported by linear \cite{Frac5,Frac11,Frac12} and nonlinear \cite{Frac-NL} periodic potentials which refer, respectively, to optical lattice and nonlinear lattice as described below.

It is well-known that the competing cubic and quintic nonlinear terms---cubic-quintic model--- that have opposite strengths of nonlinearities (the most often used is the case with self-focusing cubic and self-defocusing quintic nonlinear terms), can help to generate and stabilize various solitons \cite{soliton-rev1,soliton-rev2,CQ1,CQ2,CQ3,CQ4,CQ5,CQ6,CQ7,CQ8}. In particular, the quintic nonlinearity may be realized in the background of optics with metal-dielectric nanocomposites by varying the proportion of silver nanoparticles suspended and the host medium \cite{Cid,Cid-AOP},  and in the context of Bose-Einstein condensates composed of a dense atom cloud by considering the influence of three-body interactions (three-body collisions on the account of s-wave low-energy atom-atom scattering) whose value and even the sign can be tuned in experiments by utilizing the commonly used Feshbach resonance technique \cite{Feshbach-review}. In one-dimensional (1D) case, the self-focusing quintic nonlinearity leads to the critical collapse of solitary waves \cite{NLus,NLus2}. In recent years, the model with competing focusing and defocusing nonlinearities has been introduced to the formation of quantum droplets, in the framework of mean-field theory with Lee-Huang-Yang corrections, in the contexts of binary  Bose-Einstein condensates with attractive interspecies interactions \cite{LHY1,LHY2}.

Another stabilization mechanism of various kinds of solitons is dominated by non-uniform media (both optical and matter-wave media). The periodic potentials such as the photonic crystals \cite{PC} and lattices \cite{WL,PL,PL2,BGS,WGA1,WGA2,WGA3,PTL1,PTL2} in terms of optics, and optical lattices \cite{soliton-rev1,BEC-OL,TPOL,EPOL,TPOL2,dgsOL} in BECs, contributed to the generation of fundamental solitons and gap ones. Notably, the solitons of latter type can be localized within the finite band gap of the underlying linear spectrum and be supported by self-defocusing nonlinearity. The non-uniform media assisted by nonlinear lattices---the periodic distribution of the strength or even the sign of nonlinearity---can give rise to different soliton families \cite{Frac-NL,soliton-rev1,soliton-rev2,CQ3,NLus,NLus2,NL1,NL2,NL3,NL4,NL5,NL6,NL7,NL8,NL9,NL10} . Particularly, the competing cubic-quintic nonlinear lattices (both the self-focusing cubic and self-defocusing quintic nonlinear terms are integrated with nonlinear lattices, with commensurate and incommensurate periods of the two lattices) were predicted to be a feasible and effective way to generate stable 2D solitons and vortices \cite{NL5}. The solitons in the models with combined linear and nonlinear lattices have been and are still being comprehensively researched in recent years \cite{LNL2da,LNL1d,LNL2d,LNL2dc}. The purely nonlinear defocusing media with spatially inhomogeneous nonlinearity whose local strength grows quickly enough from the center toward periphery in the D-dimensional coordinate, which are built on self-defocusing background and therefore do not possess critical and supercritical collapses---typical characteristics for solitons in self-focusing media, enriched the generation of various families of stable solitons and soliton composites in the self-trapping regime \cite{DF1,DF2,DF3,DF4,DF5,DF6,DF6b,DF6c,DF7,DF8,DF9,DF10,DF11,DF-FT}, such as the fundamental solitons for all (D-dimensional) space coordiantes \cite{DF1,DF2}, 1D multihump states in forms of dipole and multipole solitons \cite{DF1,DF2,DF3}, 2D bright solitary vortices carrying with an arbitrarily high topological charge \cite{DF1,DF2}, 2D localized dark solitons and vortices \cite{DF10},  multifarious 3D localized modes that are comprised of soliton gyroscopes \cite{DF6b} and skyrmions \cite{DF6c}, and very recently the flat-top solitons (in both 1D and 2D spaces) and 2D vortices \cite{DF-FT,FT-JOSAB}, to name just some of them.

Despite excellent research works on solitons in fractional Schr\"{o}dinger equation are making headway in past few years \cite{accessible1,accessible2,Frac13,Frac14,Frac5,Frac11,Frac12,Frac-NL}, the presence of solitons and their propagation properties in periodic potentials with quintic nonlinearity or in purely cubic-quintic model (without any external potential) mentioned above and combinations thereof are yet for investigating. In this article, we incorporate an external linear potential (optical lattice) into the 1D cubic-quintic or quintic-only NLFSE and examine the formation and propagation dynamics of localized gap modes. The latter model (purely quintic NLFSE) gives rise to two types of gap solitons existing as on- and off-site modes grounded on the self-focusing and self-defocusing nonlinearities, respectively. The former model (cubic-quintic NLFSE) supports stable off-site fundamental gap solitons and dipole gap modes. Assisted by linear-stability analysis and direct simulations, stability regions of all the localized gap modes are given. All the stable localized gap modes are found to match the anti-Vakhitov-Kolokolov (anti-VK) criterion \cite{VK}, $\partial P/\partial b<0$, with the total power (soliton's norm) $P$ and propagation constant $b$.

The remaining content of this article is arranged in this way. To begin with, we introduce the model and its numerical methods including linear-stability analysis and the efficient ways to search for stationary solutions and to check their dynamical properties in Sec. \ref{sec2}. Numerically found gap mode solutions in the form of cusplike gap modes are presented in Sec. \ref{sec3} and which, adhered to the quintic-only (Sec. \ref{sec3a})  and cubic-quintic (Sec. \ref{sec3b})  NLFSE models under consideration, is separated into two parts: Sec. \ref{sec3a} firstly depicts the relevant band spectrum for the linearization of the considered physical model and then investigate the existence and stability properties of gap solitons thereof under the conditions of self-defocusing and self-focusing quintic nonlinearities, with an emphasis on those gap modes lying in the first and second finite band gaps of the underlying band spectrum;  Sec. \ref{sec3b} researches on the contribution of competing cubic-quintic nonlinear terms to the stabilization of diverse localized gap modes manifesting as fundamental gap solitons and their coupled modes (dipole gap solitons), dynamical evolutions of stable and unstable gap modes of both types are displayed in a systematic form. Sec. \ref{sec4} is the peroration of this article.

\section{Theoretical model}
\label{sec2}
We consider the propagation of laser beams along the $z$ axis coordinate of the optical periodic (an optical lattice) medium with a fractional-order diffraction and cubic-quintic nonlinearities, which can be described by the above-mentioned NLFSE model for the dimensionless field amplitude $E$:
\begin{equation}
\begin{aligned}
i\frac{\partial E}{\partial z}&=\frac{1}{2}\left(-\frac{\partial^2}{\partial x^2}\right)^{\alpha/2}E +[V(x)-\gamma\left|E\right|^2+\delta\left|E\right|^4]E,
\end{aligned}
\label{NLFSE}
\end{equation}
where $E$ and $z$ stands for the field amplitude and propagation distance respectively, $V$ is the linear potential trap, $\gamma>0$ being the (cubic) self-focusing nonlinearity. It should be pointed out that $(-\partial^2/\partial x^2)^\alpha$ denotes the fractional Laplacian where $\alpha$ ($1<\alpha\leq2$) stands for the L\'{e}vy index. In particular, Eq. (\ref{NLFSE}) degenerates to the normal nonlinear Schr\"{o}dinger equation if $\alpha=2$. The linear periodic potential trap---optical lattice $V$ considered in this work is written as:
\begin{equation}
V=V_0{\rm sin}^2(x),
\end{equation}
where $V_0>0$ denotes the amplitude of optical lattice.

The substitution of $E=U~{\rm exp}(ibz)$ ($b$ is the propagation constant) in Eq. (\ref{NLFSE})  results in the following stationary equation for the stationary field amplitude $U$:

\begin{equation}
\begin{aligned}
-bU&=\frac{1}{2}\left(-\frac{\partial^2}{\partial x^2}\right)^{\alpha/2}U+[V(x)-\gamma\left|U\right|^2+\delta\left|U\right|^4]U.
\end{aligned}
\label{NLFSES}
\end{equation}
For the sake of discussion, it is necessary to consider the relationship between the soliton power $P$ and propagation constant $b$ of localized gap modes calculated by Eq. (\ref{NLFSES}), here $P$ is defined as $P=\int\left|U(x)\right|^2dx$.

On the other hand, the linear stability analysis of the so-found localized gap modes is another indispensable issue. In this article, we define the perturbed field amplitude as $U=[U(x)+p(x){\rm exp}(\lambda z)+q^*(x){\rm exp}(\lambda ^*z)]{\rm exp}(ibz)$, where $U(x)$ denotes undisturbed field amplitude, $p(x)$ and $q^*(x)$ correspond to small perturbations with eigenvalue $\lambda$. Under such perturbation, the relevant eigenvalue problem of Eq. (\ref{NLFSE}) is given by:
\begin{equation}
\left\{
\begin{aligned}
i\lambda p=&+\frac{1}{2}\left(-\frac{\partial^2}{\partial x^2}\right)^{\alpha/2}p+(b+V)p \\
&-\gamma U^2(2p+q)+\delta U^4(3p+2q),\\
i\lambda q=&-\frac{1}{2}\left(-\frac{\partial^2}{\partial x^2}\right)^{\alpha/2}q-(b+V)q \\
&+\gamma U^2(2q+p)-\delta U^4(3q+2p).
\end{aligned}
\label{LAS}
\right.
\end{equation}
It should be noted that only when all the real parts $(\lambda_{\rm R})$ of the eigenvalues $\lambda$, which are calculated by the eigenvalue equations (\ref{LAS}), are null, (that is, $\lambda_{\rm R}=0$), the perturbed solutions are stable. In this article, we employ the modified squared-operator method \cite{MSOM} and split-step Fourier method to calculate the stationary solutions of Eq. (\ref{NLFSES}) and execute numerical propagation simulations of the perturbed solutions thus found using Eq. (\ref{NLFSE}), respectively.

\begin{figure}[tbp]
\begin{center}
\includegraphics[width=1\columnwidth]{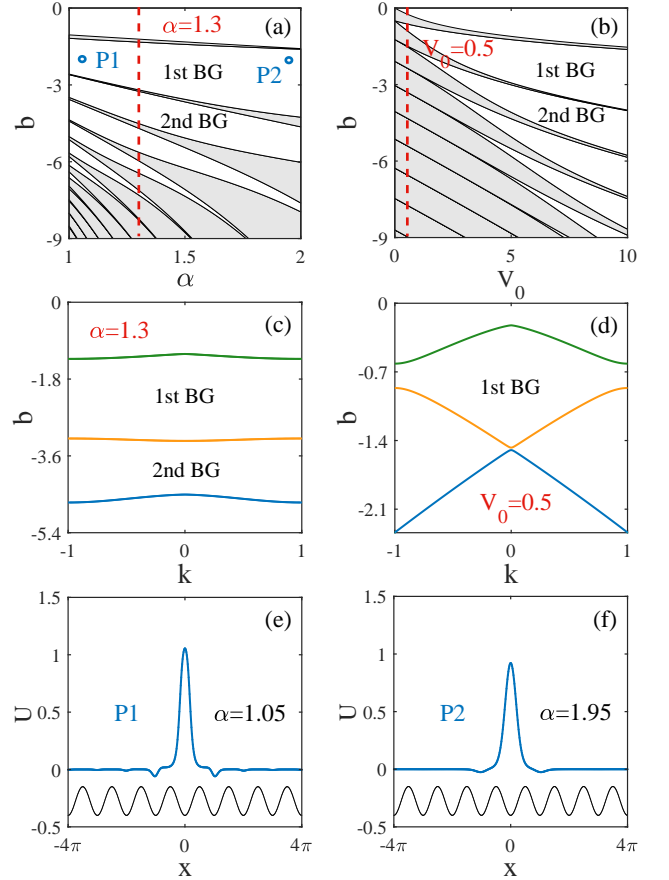}
\end{center}
\caption{(a) Band gap structure for Bloch-wave spectrum with different L\'{e}vy index $\alpha$ at $V_0=6$. The Bloch-wave spectrum of $\alpha=1.3$ is presented in Fig. \ref{fig1}(c). The profiles of the gap solitons marked by P1, P2 are displayed in Figs. \ref{fig1}(e, f) respectively. (b) Band gap structure for Bloch-wave spectrum with different $V_0$ at $\alpha=1.3$. Linear Bloch spectrum with different $V_0$ for $\alpha=1.3$: (c) at $V_0=6$; (d) at $V_0=0.5$. Profiles of off-site gap solitons under self-defocusing quintic nonlinearity ($\gamma=0$, $\delta=1$) with different $\alpha$ for $V_0=6$, $b=-2$: (e) at $\alpha=1.05$; (f) at $\alpha=1.95$.}
\label{fig1}
\end{figure}

\begin{figure}[tbp]
\begin{center}
\includegraphics[width=1\columnwidth]{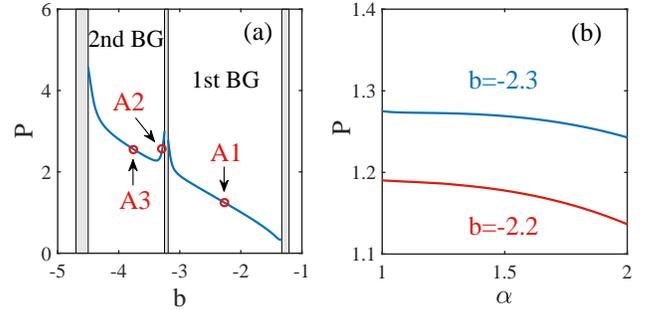}
\end{center}
\caption{(a) Soliton power $P$ versus $b$ for off-site gap solitons under self-defocusing quintic nonlinearity at $\alpha=1.3$. The profiles of the points marked by A1, A2 and A3 are displayed in Figs. \ref{fig3} (a$\sim$c) respectively. (b) Soliton power $P$ versus $\alpha$ for off-site gap solitons with different values of $b$. $\gamma=0$, $\delta=1$, $V_0=6$ for both panels.}
\label{fig2}
\end{figure}

\section{Numerical results}
\label{sec3}

We now proceed to present numerical results of different types of localized gap modes such as gap solitons and their dipole counterparts in the 1D NLFSE model (\ref{NLFSE}) within the contexts of the purely quintic nonlinearity ($\gamma=0$) and the one with competing cubic-quintic nonlinear terms; the relevant contents are therefore divided into two parts, which are referred to Sec. \ref{sec3a} and Sec. \ref{sec3b} respectively.

\subsection{Band spectrum and gap solitons in quintic nonlinearity}
\label{sec3a}

The linearization of the stationary equation  (\ref{NLFSES}), i.e., by setting the nonlinear coefficients
$\gamma=\delta=0$, leads to the linear expression of the considered physical system which, grounding on the Bloch's theorem, possesses periodic solutions known as Floquet-Bloch modes whose folding in momentum $k$ forms Bloch bands. Depicted in Figs. \ref{fig1}(a$\sim$d) show the relevant band gap structures under different physical parameters, such as at certain depth $V_0$ of the optical lattice but varying L\'{e}vy index $\alpha$, varying $V_0$ at constant $\alpha$. Basing on the  Fig. \ref{fig1}(a), multiple higher band gaps open at small $\alpha$, this is more apparent when $\alpha$ is approaching the minimum possible value 1, while most of these gaps close when increasing $\alpha$. An increase of $V_0$ at a given value of $\alpha$ (i.e., $\alpha=1.3$) brings about higher band gaps opening, as can be seen from  Fig. \ref{fig1}(b), resembling those found in \cite{Frac5}. What we are interested in is the former two finite band gaps---the first band gap and the second one, because inside both gaps where the localized modes in the forms of gap solitons and other types are usually dwell, see the case in  Fig. \ref{fig1}(c) that is focused mainly in this article. For the physical model under the condition of $\alpha=1.3$ , $V_0=0.5$ is the critical value at which the second band gap closes off completely, and the second and third Bloch bands start to hold hands, with emergence of an overlapped point [see Fig. \ref{fig1}(d)].

\begin{figure}[tbp]
\begin{center}
\includegraphics[width=1\columnwidth]{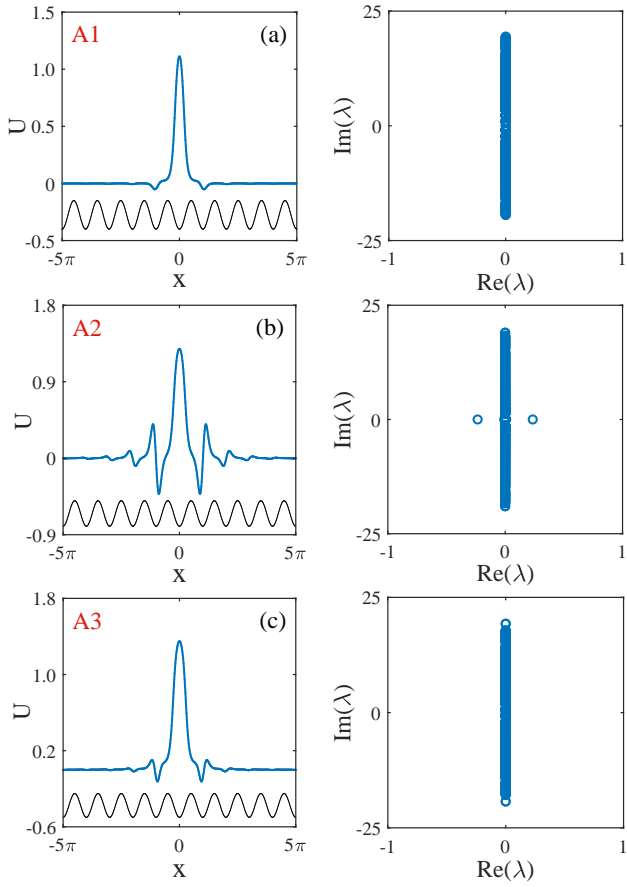}
\end{center}
\caption{Profiles and eigenvalues [given by Eq. (\ref{LAS})] of off-site gap solitons with different $b$ for quintic-only defocusing ($\gamma=0$, $\delta=1$) nonlinearity: (a) $b=-2.3$; (b) $b=-3.3$; (c) $b=-3.8$. The propagation simulations of the points A1$\sim$A3 are reported in Figs. \ref{fig5}(a$\sim$c) respectively. $V_0=6$, $\alpha=1.3$ are used in this figure and the rest of this paper.}
\label{fig3}
\end{figure}

\begin{figure}[tbp]
\begin{center}
\includegraphics[width=1\columnwidth]{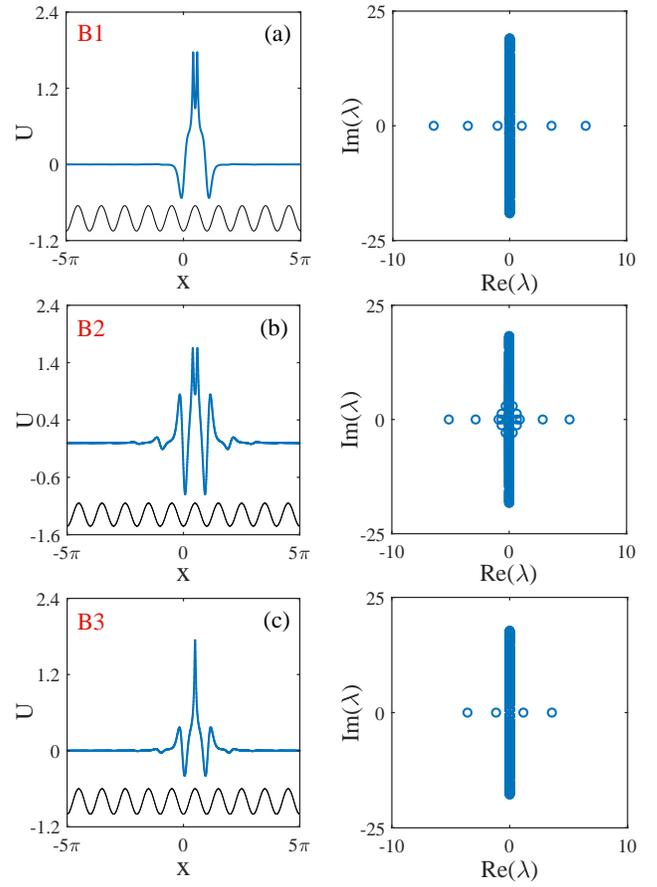}
\end{center}
\caption{Profiles and eigenvalues [given by Eq. (\ref{LAS})] of on-site gap solitons with different $b$ for quintic-only self-focusing ($\gamma=0$, $\delta=-1$) nonlinearity: (a) $b=-2.6$; (b) $b=-3.4$; (c) $b=-3.9$. Propagations of the points B1$\sim$B3 are depicted in Figs. \ref{fig5}(d$\sim$f) respectively.}
\label{fig4}
\end{figure}

Our attention for the creation of gap solitons under NLFSE model with an optical lattice is firstly restricted for the quintic-only nonlinearity as described earlier. Two characteristic examples of gap solitons stood by such model at two allowable limit values of $\alpha$, with one approaching the minimum limit $1$ another coming close to the above $2$, are portrayed respectively in  Figs. \ref{fig1}(e) and \ref{fig1}(f). The comparison research of both cases reveals that multiple modulated peaks appear to the gap soliton for the former, while such peaks constraint quickly (with perfect near-zero tails on both ends) for the latter.

The nonmonotonic relationship between power $P$ and propagation constant $b$ for the off-site gap solitons, populated in the first finite band gap and the second one of the underlying linear spectrum, in the case of quintic nonlinearity, is displayed in Fig. \ref{fig2}(a) which, with careful observation and study, suggests that the optical lattice (spatially periodic linear potential) might stabilize gap solitons in line with the anti-VK criterion. Actually, the relevant linear-stability analysis based on the numerical calculation of eigenvalue equations (\ref{LAS}) together with the subsequent direct simulations of the so-found gap solitons in dynamical equation (\ref{NLFSE}) prove that the portion of the soliton family obeying $\partial P/\partial b<0$ is linearly stable, while its counterpart with $\partial P/\partial b>0$ is linearly unstable. The two stable gap solitons (points A1 and A3 ) and one unstable case (A2), marked in the same panel and showcased in below, exemplify such argument; observably, profiles and the corresponding eigenvalues obtained from linear-stability analysis of these cusplike gap modes supported by defocusing quintic nonlinearity ($\delta=1$) are displayed in Fig. \ref{fig3} which, particularly, shows that the unstable gap soliton exhibits a higher and more modulated side peaks (highly cusplike behavior) compared to that of stable ones. In contrast, a large number of numerical computations verify that, in the context of focusing quintic nonlinearity ($\delta=-1$) , all the gap-Townes solitons (existing as the on-site localized modes) \cite{Gap-Townes}, actually no matter in where---the first and second band gaps, are completely unstable; typical examples of such modes are shown for their shapes and eigenvalues in Fig. \ref{fig4}. Actually,  finding stable fundamental gap solitons in quintic self-focusing nonlinearity is still an open issue, to our knowledge. Instead of being unstable objects, such localized gap modes can be stable solutions pinned at the first and second finite band gaps (of the underlying linear spectrum) under the cubic self-focusing nonlinearity \cite{LNL2dc}.

\begin{figure}[tbp]
\begin{center}
\includegraphics[width=1\columnwidth]{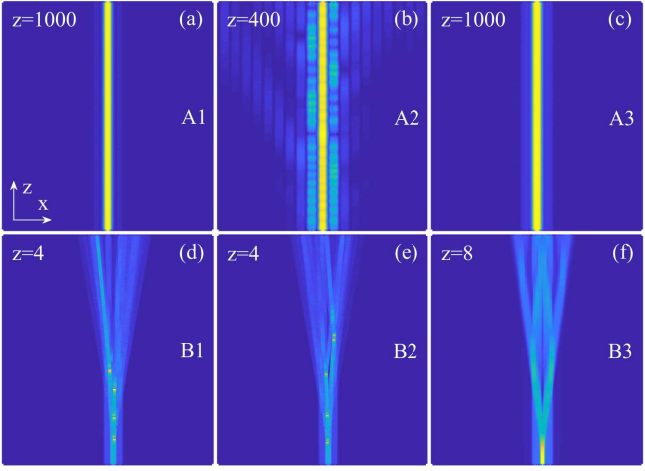}
\end{center}
\caption{Propagations of off-site gap solitons under self-defocusing quintic nonlinearity at $\delta=1$: (a) stable soliton at $b=-2.3$; (b) unstable soliton at $b=-3.3$; (c) stable soliton at $b=-3.8$. Unstable propagations of on-site gap solitons under self-focusing quintic nonlinearity at $\delta=-1$: (d) at $b=-2.6$; (e) at $b=-3.4$; (f) at $b=-3.9$. $\gamma=0$, $x\in[-30, 30]$ for all panels.}
\label{fig5}
\end{figure}

Dynamic propagation properties of the gap solitons against arbitrarily small initial perturbations, supported by the NLFSE model with an optical lattice embedded with quintic nonlinearity, are portrayed in Fig. \ref{fig5}. While stable gap solitons can propagate over a very long distance (e.g., till $z=1000$) without any distortion, the unstable localized gap modes gradually spread with emergent side peaks during the propagation and loose their coherence finally, as can be clearly seen from the figure.

\subsection{Gap solitons and dipole modes in cubic-quintic nonlinearity}
\label{sec3b}

Next, we turn to focus on how to generate gap solitons and dipole modes and what are the propagation properties of these self-trapped modes, in the NLFSE model with linear periodic potential (optical lattice) in which the saturation of nonlinear optical materials is characterized by the self-focusing cubic and self-defocusing quintic nonlinearities. For the convenience of discussion that follows, the nonlinear coefficient for the quintic term in Eqs. (\ref{NLFSE}) and (\ref{NLFSES}) is defined as $\delta=1$.

\begin{figure}[tbp]
\begin{center}
\includegraphics[width=1\columnwidth]{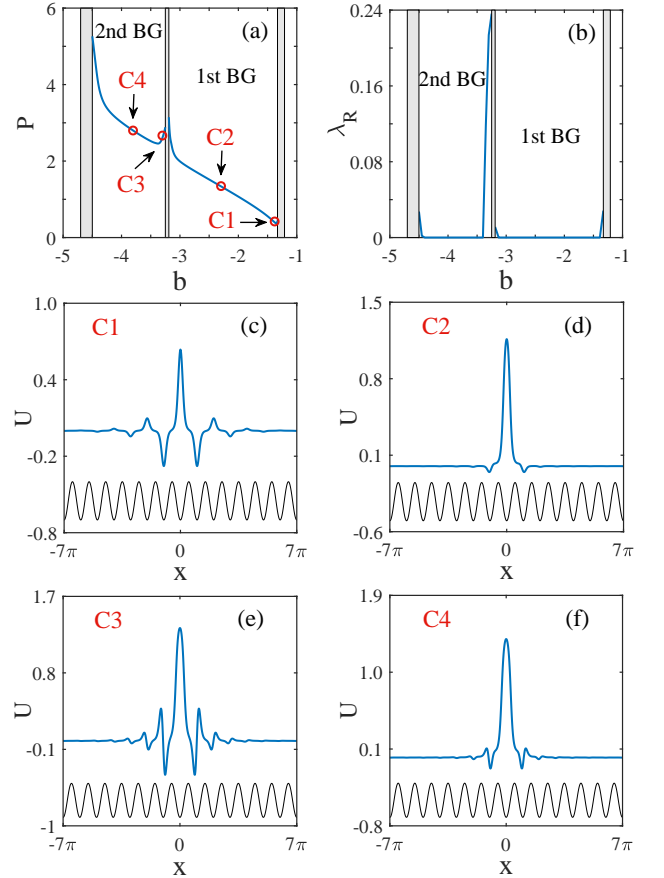}
\end{center}
\caption{(a) Soliton power $P$ and (b) maximal real part of eigenvalues ${\rm \lambda_R}$ [given by Eq. (\ref{LAS})] versus propagation constant $b$ for fundamental off-site gap solitons under competing cubic-quintic nonlinearities. Profiles of off-site gap solitons with different values of $b$: (c) at $b=-1.34$; (d) at $b=-2.3$; (e) at $b=-3.3$; (f) at $b=-3.8$.  The propagation simulations of the points C2 and C3 [whose profiles are presented in Figs. \ref{fig6}(d) and \ref{fig6}(e) respectively] are displayed in the Figs. \ref{fig9}(a) and \ref{fig9}(d), respectively. $\gamma=0.2$, $\delta=1$, for all panels.}
\label{fig6}
\end{figure}

The dependence $P(b)$  for the gap solitons and the corresponding linear-stability analysis results expressed as the dependency between  the maximal real part of eigenvalues ${\rm \lambda_R}$ [originated from Eq. (\ref{LAS})] and propagation constant $b$ are shown in Figs. \ref{fig6}(a) and \ref{fig6}(b) respectively. Both panels demonstrate once again that the gap solitons, in the context of the combination of optical lattice and competing cubic-quintic nonlinearities, exhibit the same existence and stability properties as their counterparts supported by the purely quintic nonlinear term [e.g., see Fig. \ref{fig2}(a)], that is, stable gap solitons are within the central parts of the finite band gaps (both the first and second gaps) of the underlying band structure, while unstable localized gap states are merely exist near the band edges. Furthermore, the entire family of stable gap solitons is in accordance with the abovementioned anti-VK criterion $\partial P/\partial b<0$.  The characteristic profiles of the unstable gap solitons are displayed in Figs. \ref{fig6}(c) and \ref{fig6}(e), exhibiting a cusplike envelope with highly modulated sidepeaks and covering many cells of the lattice; on contrary, stable gap modes with moderate and fewer sidepeaks are pinned at several lattice sites, according to the Figs. \ref{fig6}(d) and \ref{fig6}(f).

\begin{figure}[tbp]
\begin{center}
\includegraphics[width=1\columnwidth]{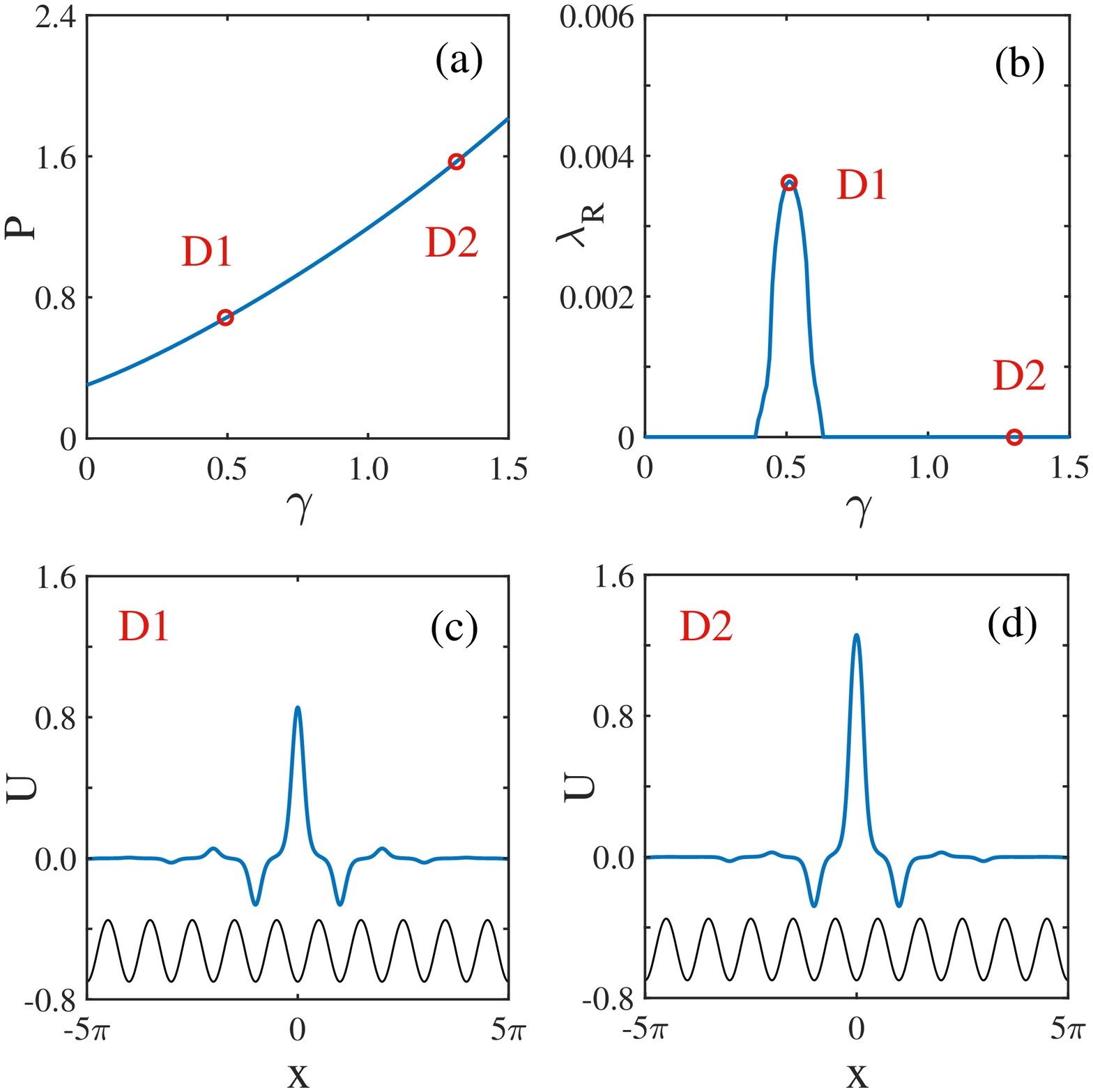}
\end{center}
\caption{(a) Soliton power $P$ and (b) maximal real part of eigenvalues ${\rm \lambda_R}$ [given by Eq. (\ref{LAS})] versus $\gamma$ for off-site fundamental gap solitons under competing cubic-quintic nonlinearities. Profiles of fundamental gap solitons with different $\gamma$: (c) at $\gamma=0.5$; (d) at $\gamma=1.3$. The propagation simulations of the points D1 and D2 [whose profiles are presented in Figs. \ref{fig7}(c) and \ref{fig7}(d) respectively] are displayed in Figs. \ref{fig9}(e) and \ref{fig9}(b), respectively. $\delta=1$, $b=-1.35$ for all panels.}
\label{fig7}
\end{figure}

The first two panels [\ref{fig7}(a) and \ref{fig7}(b)] in the Fig. \ref{fig7} display that the stability of gap solitons lying in the first finite band gap at fixing propagation constant $b=-1.35$ changes greatly under the action of self-focusing cubic nonlinearity (whose strength is denoted by $\gamma$), keep in mind that we have already set the coefficient of the quintic term as $\delta=1$; this finding can also be corroborated by direct simulation of their perturbed propagations along the axis coordinate $z$,  the relevant evolutions are shown below in the Fig. \ref{fig9} and from which, we can see that the stable gap solitons' shapes can be well conserved over long propagation distance, while unstable ones loose their coherence and start to attenuate in the propagation process and will eventually evolve into radiating waves [according to the Fig. \ref{fig9}(e)]. It is seen that the power $P$ is linear with an increase of $\gamma$, and the unstable gap solitons at the given propagation constant ($b=-1.35$) are only confined in limit region of Kerr nonlinear strength represented by $\gamma \in [0.39, 0.67]$ [see the Fig. \ref{fig7}(b)]. Representatives for unstable gap mode and stable one are respectively depicted in Figs. \ref{fig7}(c) and \ref{fig7}(d), exhibiting a cusplike behavior in profiles in accordance with their counterparts supported by quintic-only nonlinearity [cf. Figs. \ref{fig3} and \ref{fig4}].

\begin{figure}[tbp]
\begin{center}
\includegraphics[width=1\columnwidth]{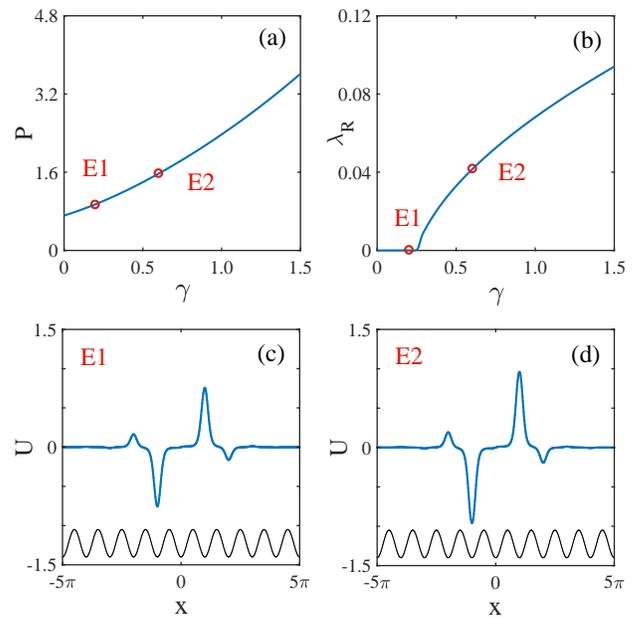}
\end{center}
\caption{(a) Soliton power $P$ and (b) maximal real part of eigenvalues ${\rm \lambda_R}$ [given by Eq. (\ref{LAS})] versus $\gamma$ for off-site dipole gap solitons under competing cubic-quintic nonlinearities. Profiles of dipole gap solitons with different $\gamma$: (c) at $\gamma=0.2$; (d) at $\gamma=0.6$. The propagation simulations of the points E1 and E2 [whose profiles are presented in Figs. \ref{fig8}(c) and \ref{fig8}(d) respectively] are displayed in Figs. \ref{fig9}(c) and \ref{fig9}(f), respectively. $\delta=1$, $b=-1.4$ for all panels.}
\label{fig8}
\end{figure}

\begin{figure}[htbp]
\begin{center}
\includegraphics[width=1\columnwidth]{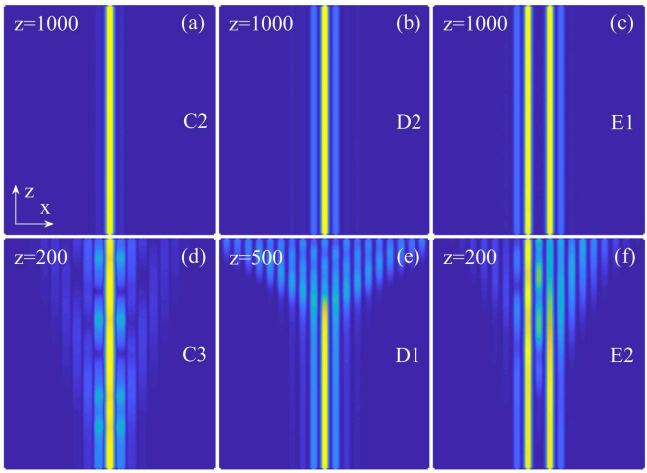}
\end{center}
\caption{Stable propagations of gap solitons supported by competing cubic-quintic nonlinearities: (a) fundamental soliton at $b=-2.3$, $\gamma=0.2$; (b) fundamental soliton at $b=-1.35$, $\gamma=1.3$; (c) dipole soliton at $b=-1.4$, $\gamma=0.2$. Unstable propagations of gap solitons: (d) fundamental soliton at $b=-3.3$, $\gamma=0.2$; (e) fundamental soliton at $b=-1.35$, $\gamma=0.5$; (f) dipole soliton at $b=-1.4$, $\gamma=0.6$. $\delta=1$, $x\in[-30, 30]$ for all panels.}
\label{fig9}
\end{figure}

In addition to the fundamental gap solitons presented above, the NLFSE model also supports a vast variety of bound solitons appearing as dipole gap modes with a spacing interval ($\delta_x$) between the two gap solitons equalling to twice the period of the optical lattice ($P_{latt}=\pi$), $\delta_x=2P_{latt}=2\pi$; two typical examples of such localized modes are illustrated in Figs. \ref{fig8}(c) and \ref{fig8}(d). It is necessary to point out that such dipole soliton family, arranged in the way with spacing interval commensurate to the period of optical lattice ($\delta_x=P_{latt}=\pi$), can not be constructed as stationary solutions, corroborated by our systematic study of the linear stability properties of the so-found solutions and the subsequent direct simulations. Different from the case for the gap solitons that are stable objects in a wide range of $\gamma$ [cf. Figs. \ref{fig7}(a) and \ref{fig7}(b)], the dipole gap solitons at determinate propagation constant ($b=-1.4$ in the first finite band gap) and quintic nonlinear coefficient ($\delta=1$) are stable modes only if the Kerr coefficient is relatively small, e.g., $\gamma \leq 0.24$ in the Fig. \ref{fig8}(b). Stable and unstable evolutions of the predicted dipole gap solitons are independently displayed in Figs. \ref{fig9}(c) and \ref{fig9}(f), and particularly, the latter panel shows the collapse of unstable dipole mode caused by internal attenuation and oscillation, and not by the reflection from both ends since the calculated size $x$ has been truncated to a very long length in our numerical experiments, indicating that the instability originates from inner change (the intrinsic property) of the bound state.

\section{Conclusion}
\label{sec4}

In this article, we have investigated the existence and propagation properties of localized gap modes, including both the fundamental gap solitons (single peak modes) and their dipole ones, in the framework of a newly introduced model---nonlinear fractional Schr\"{o}dinger equation (NLFSE) with an external periodic potential (optical lattice or photonic crystal) embedded with purely quintic nonlinearity or cubic-quintic nonlinearities. The model plays an important role in creating and stabilizing various solitary waves, owning to an effect of fractional-order diffraction thereof (denoted by L\'{e}vy index $\alpha$ ) and its interaction with the material nonlinearity is new, interesting and intriguing, compared to the standard model (usual nonlinear Schr\"{o}dinger equation with constant diffraction $\alpha=2$). In the context of purely quintic nonlinearity, the fundamental gap solitons, which lies within both the first and second band gaps of the underlying linear spectrum, are demonstrated to be unstable localized states (gap-Townes solitons) when the nonlinearity is self-focusing, and be stable profiles with broad stability regions (inside the first two finite band gaps) under the self-defocusing effect. In the competing (self-focusing) cubic-( self-defocusing) quintic nonlinearities, there are two kinds of cusplike localized states which can be viewed as fundamental gap solitons and dipole ones, and notably, the stability region of their existence is much broader for the former than that for the latter. The existence of stability and instability regions of both localized states under the two types of nonlinearities is identified by linear-stability analysis and direct simulations.

Just like the investigation of dispersion management methods in optical fiber systems has led to the formation of dispersion-managed solitons which in turn improved the performance of the optical communications, the manipulation of light diffraction (fractional diffraction) property would definitely benefit the control of light propagations, especially in nonlinear regime. On the basis of Longhi's proposal to Fractional Schr\"{o}dinger equation in optics \cite{Frac-OL}, we envision that the proposed model and the corresponding localized modes acquired here might be implemented in optics experiment with a replacement of the fractional quantum harmonic oscillator by a periodic medium whose nonlinear landscapes feature a competing cubic-quintic form.

\begin{acknowledgements}
This work was supported, in part, by the the Natural Science Foundation of China (project Nos. 61690222, 61690224), and by the Youth Innovation Promotion Association of the Chinese Academy of Sciences (project No. 2016357).
\end{acknowledgements}

\section*{Conflict of interest}
\label{sec5}
The authors declare that they have no conflict of interest.


\begin{thebibliography}{}
\bibitem{Bender-PT1} Bender, C. M. and Boettcher, S., Real Spectra in Non-Hermitian Hamiltonians Having $\mathcal{PT}$ Symmetry, Phys. Rev. Lett. \textbf{80}, 5243 (1998).
\bibitem{Bender-PT2} Bender, C. M., Brody, D. C., and Jones, H. F., Complex Extension of Quantum Mechanics, Phys. Rev. Lett. \textbf{89}, 270401 (2002).
\bibitem{Bender-PT3} Bender, C. M., Brody, D. C., Jones, H. F., and Meister, B. K., Faster than Hermitian Quantum Mechanics, Phys. Rev. Lett. \textbf{98}, 040403 (2007).
\bibitem{PT-our} Zeng, J. and Lan, Y., Two-dimensional solitons in $\mathcal{PT}$ linear lattice potentials, Phys. Rev. E \textbf{85}, 047601 (2012).
\bibitem{PT-rev1} El-Ganainy, R., Makris, K. G., Khajavikhan, M., Musslimani, Z. H., Rotter, S., and Christodoulides, D. N., Non-Hermitian physics and PT symmetry, Nature Phys. \textbf{14}, 11 (2018).
\bibitem{PT-rev2} Konotop, V. V., Yang, J., and Zezyulin, D. A. Nonlinear waves in $\mathcal{PT}$-symmetric systems, Rev. Mod. Phys. \textbf{81}, 013624 (2016).
\bibitem{PT-rev3} Suchkov, S. V., Sukhorukov, A. A., Huang, J., Dmitriev, S. V., Lee, C., and Kivshar, Y. S., Nonlinear switching and solitons in PT-symmetric photonic systems, Laser Photonics Rev. \textbf{10}, 177 (2016).

\bibitem{Lask1} Laskin, N., Fractional quantum mechanics and L\'{e}vy path integrals, Phys. Lett. A \textbf{268}, 298-305 (2000).
\bibitem{Lask2} Laskin, N., Fractional quantum mechanics, Phys. Rev. E \textbf{62}, 3135 (2000).
\bibitem{Lask3} Laskin, N., Fractional Schr\"{o}dinger equation, Phys. Rev. E \textbf{66}, 056108 (2002).
\bibitem{Frac1} Herrmann, R., \emph{Fractional Calculus: An Introduction for Physicists}, (World Scientific, Singapore, 2011).
\bibitem{Frac-condensed} Stickler, B. A., Potential condensed-matter realization of space-fractional quantum mechanics: The one-dimensional L\'{e}vy crystal, Phys. Rev. E \textbf{88}, 012120 (2013).

\bibitem{Frac-OL} Longhi, S., Fractional Schr\"{o}dinger equation in optics, Opt. Lett. \textbf{40}, 1117 (2015).

\bibitem{Frac2} Zhang, Y., Liu, X., Beli\'{c}, M. R., Zhong, W., Zhang, Y., and Xiao, M., Propagation dynamics of a light beam in a fractional Schr\"{o}dinger equation, Phys. Rev. Lett. \textbf{115}, 180403 (2015).
\bibitem{Frac6} Zhang, Y., Zhong, H., Beli\'{c}, M. R., Ahmed, N., Zhang, Y., and Xiao, M., Diffraction-free beams in fractional Schr\"{o}dinger equation, Sci. Rep. \textbf{6}, 23645 (2016).
\bibitem{Frac3} Zhang, Y., Zhong, H., Beli\'{c}, M. R., Zhu, Y., Zhong, W., Zhang, Y., Christodoulides, D. N. and Xiao, M., $\mathcal{PT}$ symmetry in a fractional Schr\"{o}dinger equation, Laser Photonics Rev. \textbf{10}, 526 (2016).
\bibitem{Frac7} Zhang, L., Li, C., Zhong, H., Xu, C., Lei, D., Li, Y., and Fan, D., Propagation dynamics of super-Gaussian beams in fractional Schr\"{o}dinger equation: From linear to nonlinear regimes, Opt. Express \textbf{24}, 14406 (2016).
\bibitem{accessible1} Zhong, W. P., Beli\'{c}, M. R., Malomed, B. A., Zhang, Y., and Huang, T., Spatiotemporal accessible solitons in fractional dimensions, Phys. Rev. E \textbf{94}, 012216 (2016).
\bibitem{accessible2} Zhong, W. P., Beli\'{c}, M. R., and Zhang, Y., Accessible solitons of fractional dimension, Ann. Phys.  \textbf{368}, 110 (2016).
\bibitem{Frac8} Zhang, Y., Wang, R., Zhong, H., Zhang, J., Beli\'{c}, M. R., and Zhang, Y., Optical Bloch oscillation and Zener tunneling in the fractional Schr\"{o}dinger equation, Sci. Rep. \textbf{7}, 17872 (2017).
\bibitem{Frac9} Huang, C. and Dong, L., Beam propagation management in a fractional Shr\"{o}dinger equation, Sci. Rep. \textbf{7}, 5442 (2017).
\bibitem{Frac10} Zhang, L., He, Z., Conti, C., Wang, Z., Hu, Y., Lei, D., Li, Y., and Fan, D., Modulational instability in fractional nonlinear Schr\"{o}dinger equation, Commun. Nonlinear Sci. Numer. Simulat. \textbf{48}, 531 (2017).

 \bibitem{Frac13} Chen, M., Zeng, S., Lu, D., Hu, W., and Guo, Q., Optical solitons, self-focusing, and wave collapse in a space-fractional Schr\"{o}dinger equation with a Kerr-type nonlinearity, Phys. Rev. E \textbf{98}, 022211 (2018).
\bibitem{Frac14} Chen, M., Guo, Q., Lu, D., and Hu, W., Variational approach for breathers in a nonlinear fractional Schr\"{o}dinger equation, Commun. Nonlinear Sci. Numer. Simulat. \textbf{71}, 73 (2019).

\bibitem{Frac5} Huang, C. and Dong, L., Gap solitons in the nonlinear fractional Schr\"{o}dinger equation with an optical lattice, Opt. Lett. \textbf{41}, 5636 (2016).
\bibitem{Frac11} Dong, L. and Huang, C., Double-hump solitons in fractional dimensions with a $\mathcal{PT}$-symmetric potential, Opt. Express \textbf{26}, 10509 (2018).
 \bibitem{Frac12} Yao, X. and Liu, X., Off-site and on-site vortex solitons in space-fractional photonic lattices, Opt. Lett. \textbf{43}, 5749 (2018).

\bibitem{Frac-NL} Zeng, L. and Zeng, J., One-dimensional solitons in fractional Schr\"{o}dinger equation with a spatially periodical modulated nonlinearity: nonlinear lattice, Opt. Lett. \textbf{44}, 2661 (2019).

\bibitem{soliton-rev1} Kartashov, Y. V., Malomed, B. A., and Torner, L., Solitons in nonlinear lattices, Rev. Mod. Phys. \textbf{83}, 247 (2011).
\bibitem{soliton-rev2} Kartashov, Y. V., Astrakharchik, G. E., Malomed, B. A., and Torner, L., Frontiers in multidimensional self-trapping of nonlinear fields and matter, Nat. Rev. Phys. \textbf{1}, 185 (2019).

\bibitem{CQ1} Triki, H., Porsezian, K., Dinda, P. T., and Grelu, P., Dark spatial solitary waves in a cubic-quintic-septimal nonlinear medium, Phys. Rev. A, \textbf{95}, 023837 (2017).
\bibitem{CQ2} Cisternas, J., Descalzi, O., Albers, T., and Radons, G., Anomalous Diffusion of Dissipative Solitons in the Cubic-Quintic Complex Ginzburg-Landau Equation in Two Spatial Dimensions, Phys. Rev. Lett. \textbf{116}, 203901, (2016).
\bibitem{CQ3} Gao, X. and Zeng, J., Two-dimensional matter-wave solitons and vortices in competing cubic-quintic nonlinear lattices, Front. Phys. \textbf{13}, 130501 (2018).
\bibitem{CQ4} Zegadlo, K. B., Wasak, T., Malomed, B. A., Karpierz, M. A., and Trippenbach, M., Stabilization of solitons under competing nonlinearities by external potentials, Chaos, \textbf{24}, 043136 (2014).
\bibitem{CQ5} Burlak, G. and Malomed, B. A., Interactions of three-dimensional solitons in the cubic-quintic model, Chaos, \textbf{28}, 063121 (2018).
\bibitem{CQ6} Desyatnikov, A., Maimistov, A., and Malomed, B., Three-dimensional spinning solitons in dispersive media with the cubic- quintic nonlinearity, Phys. Rev. E \textbf{61}, 3107 (2000).
\bibitem{CQ7} Paredes, A., Feijoo, D., and Michinel, H., Coherent cavitation in the liquid of light, Phys. Rev. Lett. \textbf{112}, 173901 (2014).
\bibitem{CQ8} Falc\~{a}o-Filho, E. L., Ara\'{u}jo, C. B. de, Boudebs, G., Leblond, H., and Skarka, V., Robust Two-Dimensional Spatial Solitons in Liquid Carbon Disulfide, Phys. Rev. Lett. \textbf{110}, 013901 (2013).

\bibitem{Cid} Reyna, A. S. and Ara\'{u}jo, C. B. de, Nonlinearity management of photonic composites and observation of spatial-modulation instability due to quintic nonlinearity, Phys. Rev. A \textbf{89}, 063803 (2014).
\bibitem{Cid-AOP} Reyna, A. S. and Ara\'{u}jo, C. B. de, High-order optical nonlinearities in plasmonic nanocomposites¡ªa review, Adv. Opt. Photon. \textbf{9}, 720 (2017).

\bibitem{Feshbach-review} Chin, C., Grimm, R., Julienne, P., and Tsienga, E., Feshbach resonances in ultracold gases, Rev. Mod. Phys. \textbf{82}, 1225 (2010).

\bibitem{NLus} Zeng, J. and Malomed, B. A., Stabilization of one-dimensional solitons against the critical collapse by quintic nonlinear lattices, Phys. Rev. A \textbf{85}, 023824 (2012).
\bibitem{NLus2} Shi, J., Zeng, J., and Malomed, B. A., Suppression of the critical collapse for one-dimensional solitons by saturable quintic nonlinear lattices," Chaos \textbf{28}, 075501 (2018).
\bibitem{LHY1} Petrov, D. S., Quantum Mechanical Stabilization of a Collapsing Bose-Bose Mixture, Phys. Rev. Lett. \textbf{115}, 155302 (2015).
\bibitem{LHY2} Petrov, D. S. and Astrakharchik, G. E., Ultradilute Low-Dimensional Liquids, Phys. Rev. Lett. \textbf{117}, 100401 (2016).

\bibitem{PC} Joannopoulos, J. D., Johnson, S. G., Winn, J. N., and Meade, R. D., \emph{Photonic Crystals: Molding the Flow of Light}, (Princeton University Press: Princeton, 2008).

\bibitem{WL} Christodoulides, D. N., Lederer, F., and Silberberg, Y., Discretizing light behaviour in linear and nonlinear waveguide lattices, Nature \textbf{424}, 817 (2003).
\bibitem{PL} Garanovich, I. L., Longhi, S., Sukhorukova, A. A., and Kivshar, Y. S., Light propagation and localization in modulated photonic lattices and waveguides, \emph{Phys. Rep.} \textbf{518}, 1 (2012).
\bibitem{PL2} Chen, Z., Segev, M., and Christodoulides, D. N., Optical spatial solitons: historical
overview and recent advances, Rep. Prog. Phys. \textbf{75}, 086401 (2012).

\bibitem{BGS} Eggleton, B. J., Slusher, R. E., Sterke, C. M. de, Krug, P. A., and Sipe, J. E., Bragg grating solitons, Phys. Rev. Lett. \textbf{76}, 1627 (1996).
\bibitem{WGA1} Mandelik, D., Morandotti, R., Aitchison, J. S., and Silberberg, Y., Gap solitons in waveguide arrays, Phys. Rev. Lett. \textbf{92}, 093904 (2004).
\bibitem{WGA2} Kartashov, Y. V., Vysloukh, V. A., and Torner, L., Surface gap solitons, Phys. Rev. Lett. \textbf{96}, 073901 (2006).
\bibitem{WGA3} Szameit, A., Kartashov, Y. V., Dreisow, F., Pertsch, T., Nolte, S., T\"{u}nnermann, A., and Torner, L., Observation of two-dimensional surface solitons in asymmetric waveguide arrays, Phys. Rev. Lett. \textbf{98}, 173903 (2007).

\bibitem{PTL1} Peleg, O., Bartal, G., Freedman, B., Manela, O., Segev, M., and Christodoulides, D. N., Conical diffraction and gap solitons in honeycomb photonic lattices, Phys. Rev. Lett. \textbf{98}, 103901 (2007).
\bibitem{PTL2} Fleischer, J. W., Segev, M., Efremidis, N. K., and Christodoulides, D. N., Observation of two-dimensional discrete solitons in optically induced nonlinear photonic lattices, Nature \textbf{422}, 147 (2003).

\bibitem{BEC-OL} Baizakov, B. B., Malomed, B. A., and Salerno, M., Multidimensional solitons in periodic potentials, Europhys. Lett. \textbf{63}, 642 (2003).
\bibitem{TPOL} Brazhnyi, V. A. and Konotop, V. V., Theory of nonlinear matter waves in optical lattices, Mod. Phys. Lett. B, \textbf{18}, 627 (2004).
\bibitem{EPOL} Eiermann, B., Anker, Th., Albiez, M., Taglieber, M., Treutlein, P., Marzlin, K.-P., and Oberthaler, M. K., Bright Bose-Einstein Gap Solitons of Atoms with Repulsive Interaction, Phys. Rev. Lett. \textbf{92}, 230401 (2004).
\bibitem{TPOL2} Morsch, O. and Oberthaler, M., Dynamics of Bose-Einstein condensates in optical lattices, Rev. Mod. Phys. \textbf{78}, 179 (2006).

\bibitem{dgsOL} Zeng, L. and J. Zeng., Gap-type dark localized modes in a Bose-Einstein condensate with optical lattices, Adv. Photonics \textbf{1}, 046004 (2019).

\bibitem{NL1} Sakaguchi, H. and Malomed, B. A., Matter-wave solitons in nonlinear optical lattices, Phys. Rev. E \textbf{72}, 046610 (2005).
\bibitem{NL2} Theocharis, G., Schmelcher, P., Kevrekidis, P. G., and Frantzeskakis, D. J., Matter-wave solitons of collisionally inhomogeneous condensates, Phys. Rev. A \textbf{72}, 033614 (2005).
\bibitem{NL3} Sivan, Y., Fibich, G., and Weinstein, M. I., Waves in nonlinear lattices: ultrashort optical pulses and Bose-Einstein condensates, Phys. Rev. Lett. \textbf{97}, 193902 (2006).
\bibitem{NL4} Belmonte-Beitia, J., P\'{e}rez-Garc\'{i}a, V. M., Vekslerchik, V., and Torres, P. J., Lie symmetries and solitons in nonlinear systems with spatially inhomogeneous nonlinearities, Phys. Rev. Lett. \textbf{98}, 064102 (2007).
\bibitem{NL5} Kartashov, Y. V., Vysloukh, V. A., and Torner, L., Soliton modes, stability, and drift in optical lattices with spatially modulated nonlinearity, Opt. Lett., \textbf{33}, 1747 (2008).
\bibitem{NL6} Kartashov, Y. V., Malomed, B. A., Vysloukh, V. A., and Torner, L., Vector solitons in nonlinear lattices, Opt. Lett., \textbf{34}, 3625 (2009).
\bibitem{NL7} Abdullaev, F. Kh., Gammal, A., Salerno, M., and Tomio, L., Localized modes of binary mixtures of Bose-Einstein condensates in nonlinear optical lattices, Phys. Rev. A, \textbf{77}, 023615 (2008).

\bibitem{NL8} Lebedev, M. E., Alfimov, G. L., and Malomed, B. A., Stable dipole solitons and soliton complexes in the nonlinear Schr\"{o}dinger equation with periodically modulated nonlinearity, Chaos, \textbf{26}, 073110 (2016).
\bibitem{NL9} Wen, Z. and Yan, Z., Solitons and their stability in the nonlocal nonlinear Schr\"{o}dinger equation with $\mathcal{PT}$-symmetric potentials, Chaos, \textbf{27}, 053105 (2017).
\bibitem{NL10} Zezyulin, D. A. and Konotop, V. V., Solitons in a Hamiltonian $\mathcal{PT}$-symmetric coupler, J. Phys. A: Math. Theor., \textbf{51}, 015206 (2018).

\bibitem{LNL2da} Kartashov, Y. V., Vysloukh, V. A., and Torner, L., Power-dependent shaping of vortex solitons in
optical lattices with spatially modulated nonlinear refractive index, Opt. Lett., \textbf{33}, 2173 (2008).
\bibitem{LNL1d} Sakaguchi, H. and Malomed, B. A., Solitons in combined linear and nonlinear lattice potentials, Phys. Rev. A \textbf{81}, 013624 (2010).
\bibitem{LNL2d} Zeng, J. and Malomed, B. A., Two-dimensional solitons and vortices in media with incommensurate linear and nonlinear lattice potentials, Phys. Scr. \textbf{T149}, 014035 (2012).
\bibitem{LNL2dc} Shi, J. and Zeng, J.,  Self-trapped spatially localized states in combined linear-nonlinear periodic potentials, Frontiers of Physics (accepted).

\bibitem{DF1} Borovkova, O. V., Kartashov, Y. V., Torner, L., and Malomed, B. A., Bright solitons from defocusing nonlinearities, Phys. Rev. E \textbf{84}, 035602(R) (2011).
\bibitem{DF2} Borovkova, O. V., Kartashov, Y. V., Malomed, B. A., and Torner, L., Algebraic bright and vortex solitons in defocusing media, Opt. Lett. \textbf{36}, 3088 (2011).
\bibitem{DF3} Zeng, J. and Malomed, B. A., Bright solitons in defocusing media with spatial modulation of the quintic nonlinearity, Phys. Rev. E \textbf{86}, 036607 (2012).
\bibitem{DF4} Kartashov, Y. V., Lobanov, V. E., Malomed, B. A., and Torner, L., Asymmetric solitons and domain walls supported by inhomogeneous defocusing nonlinearity, Opt. Lett. \textbf{37}, 5000 (2012).
\bibitem{DF5} Young-S, L. E., Salasnich, L., and Malomed, B. A., Self-trapping of Fermi and Bose gases under spatially modulated repulsive nonlinearity and transverse confinement, Phys. Rev. A \textbf{87}, 043603 (2013).
\bibitem{DF6} Cardoso, W. B., Zeng, J., Avelar, A. T., Bazeia, D., and Malomed, B. A., Bright solitons from the nonpolynomial Schr\"{o}dinger equation with inhomogeneous defocusing nonlinearities, Phys. Rev. E \textbf{88}, 025201 (2013).
\bibitem{DF6b} Driben, R., Kartashov, Y. V., Malomed, B. A., Meier, T., and Torner, L., Soliton Gyroscopes in Media with Spatially Growing Repulsive Nonlinearity, Phys. Rev. Lett. \textbf{112}, 020404 (2014).
\bibitem{DF6c} Kartashov, Y. V., Malomed, B. A., Shnir, Y., and Torner, L., Twisted Toroidal Vortex Solitons in Inhomogeneous Media with Repulsive Nonlinearity, Phys. Rev. Lett. \textbf{113}, 264101 (2014).
\bibitem{DF7} Driben, R., Kartashov, Y. V., Malomed, B. A., Meier, T., and Torner, L., Three-dimensional hybrid vortex solitons, New J. Phys. \textbf{16}, 063035 (2014).
\bibitem{DF8} Kevrekidis, P. G., Malomed, B. A., Saxena, A., Bishop, A. R., and Frantzeskakis, D. J., Solitons and vortices in two-dimensional discrete nonlinear Schr\"{o}dinger systems with spatially modulated nonlinearity, Phys. Rev. E \textbf{91}, 043201 (2015).
\bibitem{DF9} Driben, R., Dror, N., Malomed, B. A., and Meier, T., Multipoles and vortex multiplets in multidimensional media with inhomogeneous defocusing nonlinearity, New J. Phys. \textbf{17}, 083043 (2015).
\bibitem{DF10} Zeng, J. and Malomed, B. A., Localized dark solitons and vortices in defocusing media with spatially inhomogeneous nonlinearity, Phys. Rev. E \textbf{95}, 052214 (2017).
\bibitem{DF11} Huang, C., Ye, Y., Liu, S., He, H., Pang, W., Malomed, B. A., and Li, Y., Excited states of two-dimensional solitons supported by spin-orbit coupling and field-induced dipole-dipole repulsion, Phys. Rev. A \textbf{97}, 013636 (2018).
\bibitem{DF-FT} Zeng, L., Zeng, J., Kartashov, Y. V., and Malomed, B. A., Purely Kerr nonlinear model admitting flat-top solitons, Opt. Lett. \textbf{44}, 1206 (2019).
\bibitem{FT-JOSAB} Zeng, L. and Zeng, J.,  Gaussian-like and flat-top solitons of atoms with spatially modulated repulsive interactions, J. Opt. Soc. Am. B \textbf{36}, 002278 (2019).
\bibitem{VK} Vakhitov, M. and Kolokolov, A., Stationary solutions of the wave equation in a medium with nonlinearity saturation," Radiophys. Quantum Electron. \textbf{16}, 783 (1973).
\bibitem{MSOM} Yang, J., \emph{Nonlinear Waves in Integrable and Nonintegrable Systems}, (SIAM: Philadelphia, 2010).
\bibitem{Gap-Townes} Abdullaev, F. Kh. and Salerno, M., Gap-Townes solitons and localized excitations in low-dimensional Bose-Einstein condensates in optical lattices, Phys. Rev. A, \textbf{72}, 033617 (2005).

\end{thebibliography}
\end{document}